\expandafter\def\csname ver@fixltx2e.sty\endcsname{}
\documentclass[review ]{elsarticle}

\let\today\relax
\makeatletter
\def\ps@pprintTitle{%
    \let\@oddhead\@empty
    \let\@evenhead\@empty
    \def\@oddfoot{\footnotesize\itshape
         {Paper accepted at Information Processing \& Management on October 29, 2019.} \hfill\today}%
    \let\@evenfoot\@oddfoot
    }
\makeatother

\usepackage{lineno,hyperref}
\usepackage{soul}
\usepackage{enumitem} 
\usepackage[flushleft]{threeparttable}

\usepackage[font={bf,rm,rm}]{caption} 

\usepackage{multirow} 
\usepackage{booktabs} 
\usepackage{amssymb} 
\usepackage{latexsym}
\usepackage{dblfloatfix}    
\usepackage[usenames,dvipsnames]{xcolor}
\usepackage{pifont}
\usepackage{calc} 
\usepackage{dialogue}
\usepackage{adjustbox}
\usepackage{float}

\definecolor{tealblue}{rgb}{0.21, 0.46, 0.53}
\definecolor{wildstrawberry}{rgb}{1.0, 0.26, 0.64}
\definecolor{ao(english)}{rgb}{0.0, 0.5, 0.0}
\definecolor{cadetblue}{rgb}{0.37, 0.62, 0.63}
\definecolor{lavenderindigo}{rgb}{0.58, 0.34, 0.92}
\definecolor{fielddrab}{rgb}{0.42, 0.33, 0.12}

\newcommand{\ignore}[1]{}

\newcommand{\change}[1]{\textcolor{black}{#1}}

\modulolinenumbers[5]

\journal{Journal of Information Processing \& Management}

\begin{document}

\begin{frontmatter}

\title{Towards a Model for Spoken Conversational Search}

\author[rmit]{Johanne R.~Trippas\corref{mycorrespondingauthor}}
\ead{johanne.trippas@rmit.edu.au}

\author[rmit]{Damiano Spina}
\ead{damiano.spina@rmit.edu.au}

\author[msr]{Paul Thomas}
\ead{pathom@microsoft.com}

\author[rmit]{Mark Sanderson}
\ead{mark.sanderson@rmit.edu.au}

\author[tsukuba]{Hideo Joho}
\ead{hideo@slis.tsukuba.ac.jp}

\author[rmit]{Lawrence Cavedon}
\ead{lawrence.cavedon@rmit.edu.au}

\address[rmit]{RMIT University, Melbourne, Australia}
\address[msr]{Microsoft, Canberra, Australia}
\address[tsukuba]{University of Tsukuba, Tsukuba, Japan}
\cortext[mycorrespondingauthor]{Corresponding author.}

\begin{abstract}
    Conversation is the natural mode for information exchange in daily life, a spoken conversational interaction for search input and output is a logical format for information seeking.
    However, the conceptualisation of user--system interactions or information exchange in \change{spoken conversational search} (SCS) has not been explored.
    The first step in conceptualising SCS is to understand the conversational moves used in an audio-only communication channel for search. This paper explores conversational actions for the task of search.
    We define a qualitative methodology for creating conversational datasets, propose analysis protocols, and develop the \textit{SCSdata}. Furthermore, we use the SCSdata to create the first annotation schema for SCS: the \textit{SCoSAS}, enabling us to investigate interactivity in SCS.
    We further establish that SCS needs to incorporate interactivity and pro-activity to overcome the complexity that the information seeking process in an audio-only channel poses.
    In summary, this exploratory study unpacks the breadth of SCS. Our results highlight the need for integrating discourse in future SCS models and contributes the advancement in the formalisation of SCS models and the design of SCS systems.

\end{abstract}


\end{frontmatter}


\section{Introduction}
\label{sec:intro}
Voice-only, spoken conversational systems such as Google Home, Amazon Echo, or Apple Homepod, are becoming widely used. These systems can answer factoid questions. However, they are yet not able to engage in complex information seeking tasks where multiple turns are needed to exchange information, reformulate queries, or proactively recommend different search strategies.

\citet{trippas2018informing} suggested that existing information seeking models do not suffice for the increased interactivity, complexity, or the agency of SCS systems. To the best of our knowledge, only a few models include the system as an integral part of the search process~\cite{azzopardi2018conceptualizing,reichman1985getting, sitter1992modeling, vakulenko2019qrfa}. Recently, \citet{azzopardi2018conceptualizing} created a conceptual framework of the probable action and interaction space for conversational agents as a first step, acknowledging that their initial framework would need expansion and empirical evidence.

Understanding the communication behaviours of dialogue is crucial to SCS and many different annotation schemas have been developed~\cite{bunt2009dit++}. These schemas are classifications of dialogues and consider an utterance as an action inside the information exchange. Even though many different domain independent annotation schemas exist, such as DAMSL~\cite{allen1997draft}, these were mainly used for the creation of \change{spoken dialogue systems} (SDS) or the general understanding of dialogue. Information seeking actions were not covered in depth but instead broad categories such as ``answer" or ``info-request" are presented.
Recent work has made initial steps towards understanding such actions~\cite{radlinski2017theoretical, trippas2017people, trippas2019}, however, no complete set has been developed so far.

We create a first annotation schema for SCS: the \change{spoken conversational search annotation schema} (SCoSAS). The schema reveals the different atomic actions or utterance functions and interactions taken by prospective user and system in the stages of an information seeking process. Thematic analysis was used to identify and summarise communicative activities, strategies, and challenges~\cite{braun2013successful}. We provide a schematic overview of the possible interactions by either actor, which allows us to understand a seeker's conversational patterns~\cite{lai2009conversational}.

The analysis is based on an experimental lab study simulating a natural dialogue to understand how a user and system may interact. Since no existing SCS systems can reliably manage multiple turns, we used conversations between two people, the Seeker, the other an Intermediary. The information seeking dialogues between the two actors were filmed, transcribed, and annotated.
The annotated data is publicly available.\footnote{Transcripts together with codes/labels of the experiment are available from the corresponding author upon request.}

The contributions of this work include:
\begin{enumerate}[noitemsep]
  \item We create the first fully labelled dataset for SCS, the SCSdata;
  \item We define a multi-level annotation schema for SCS, SCoSAS, to identify the interaction choices for SCS;
  \item We create a new model based on multi-turn activities and multi-move utterances;
    \item We validate the proposed annotation schema by re-annotating the SCSdata and with annotating another conversational dataset;
    \item We suggest new design recommendations and hypotheses for SCS.
\end{enumerate}

The aim of this qualitative research is to explore SCS as a new search paradigm and seeks to understand the exhibited behaviours demonstrated in an ideal scenario for SCS. Thus, we aim to gain a deeper understanding and overview of the interaction behaviours of a group of participants through qualitative analysis. We first create a rich and detailed dataset through a \change{natural dialogue study} (NDS)~\cite{yankelovich2008using}, which we refer to as our observational study, to explore SCS interaction behaviours and to seek patterns within this data through an inductive method. \change{In particular, we use our observational study dataset to better understand the communication behaviours at first-hand instead of relying on questionnaires or self-report.}
The strengths of our qualitative analysis are that it provides an in-depth and detailed analysis to explain complex interactions during the information seeking process.

The remainder of the paper is organised as follows.
In Section~\ref{sec:relatedwork}, we present related work.  
In Section~\ref{sec:methodology}, we describe our methodological approach to create the SCSdata and analysis. In Section~\ref{sec:Results}, we present all the identified themes and sub-themes, and we then validate the coding consistency within the SCSdata. We further validate our coding schema in Section~\ref{sec:Validation of SCoSAS}.
Then, discussion, implications, and design recommendations for SCS are presented in Section~\ref{sec:discussion}, before concluding in Section~\ref{sec:conclusions}.

\section{Related Work}
\label{sec:relatedwork}

We organise previous work into four sections: Spoken Dialogue Systems (Section~\ref{subsec:Spoken Dialogue Systems}), annotating dialogues (Section~\ref{subsec:Annotating Dialogues}), interaction space in conversational search (Section~\ref{subsec:Interaction Space in Conversational Search}), and natural dialogue studies (Section~\ref{subsec:Natural Dialogue Study}).

\subsection{Spoken Dialogue Systems}
\label{subsec:Spoken Dialogue Systems}

SDS provide a platform for people to interact with computer applications such as databases with the use of spoken natural language. These systems exchange information on a turn-by-turn basis providing an interface between the user and the computer~\cite{gibbon1997handbook}.

In recent years, interest in SCS has grown, as speech technology~\cite{xiong2018microsoft} and machine learning for spoken systems~\citep{yang2018response} have developed. A range of SDS are available, from question answering to semi-conversational systems~\citep{mctear2016conversational}. Research has been devoted to task-oriented SDS which has defined search boundaries, such as travel planning or route planning, and can be developed with slot filling approaches~\citep{walker2001quantitative}. 
Thus, task-oriented dialogue systems are created on a particular closed domain. However, non-task-oriented dialogue systems or open-domain conversations such as search for SCS systems may not benefit from a rigid plan-based dialogue approach and introduce many new challenges~\citep{higashinaka2014towards}. These challenges include how to deal with the variety of user utterances and how answers or replies could be simplified or abstracted to generate appropriate system responses~\citep{sugiyama2013open}.

\subsection{Annotating Dialogues}
\label{subsec:Annotating Dialogues}

Research interest in SCS has increased the recording of spoken search interactions~\cite{thomas2017MISC, vakulenko2019qrfa}. Such records are a valuable source of data to understand how users interact and which tactics are used for driving effective search performance in this new search paradigm. Thus, this data is useful to understanding the characteristics of a search conversation to build SCS systems acting as a dialogue participant~\cite{gibbon1997handbook}.
\change{The spoken data recordings themselves need to be appropriately transcribed and ``annotated''~\cite{larson2012spoken}.} Thus, exposing the structure of the conversations by annotating the actions taken is one of the first steps towards analysing these spoken interactions~\cite{zarisheva2015dialog}.

Previously, much research has been devoted to creating annotation schemas and classifying taxonomies for dialogue and SDS~\citep{allen1997draft, bunt2009dit++, searle1969speech}. 
Annotating these dialogues has been based on the understanding that classifying utterances provides insight into the dialogue behaviour~\citep{reithinger1995utilizing}
additionally research on dialogue is often based on the assumption that dialogue acts provide a useful way of characterising dialogue behaviors in human--human dialogue, and potentially in human--computer dialogue as well~\cite{allen1997draft, belkin1995cases, bunt1999dynamic}. For example, annotated conversations can help to identify answers in texts or characterise user intents~\citep{qu2018analyzing}.

\change{Annotating dialogue transcriptions has been explored by sociologists (via {\em conversation analysis\/}, e.g., \cite{schegloff2000overlapping}) and socio-psychologists
(e.g., \cite{clark1991grounding}) for the purpose of understanding the organisation and communicative purpose of dialogue contributions. Within Computational Linguistics, annotation via {\em dialogue acts} -- which extend Searle's {\em speech acts} \cite{searle1969speech} by adding the social-communicative purpose -- has been
used to analyse dialogue transcriptions for the purpose of designing computational models of dialogue management. While Allen and Core's original DAMSL framework was designed for task-based dialogue, subsequent formulations have been designed for specific types of dialogues~\cite{allen1997draft}.}

Thus, several different annotation schemas have been proposed which cover the general speech interactions. Such schemas emphasised information seeking, such as the \change{dynamic interpretation theory} (DIT) by~\citet{bunt1999dynamic}. The DIT was based on the empirical investigation of spoken human--human information dialogues. \citeauthor{bunt1999dynamic} suggested that these information dialogues have two motivational sources, namely, to proceed in the task and to exchange communicative functions to drive the conversation~\cite{bunt1999dynamic}. He noticed that an information dialogue consisted of the expected greetings, apologies, and acknowledgements but also included information-exchange utterances such as questions, answers, checks, and confirmations. Later, \citeauthor{bunt2009dit++} developed an annotation schema called DIT++ for these information dialogues~\cite{bunt2009dit++}. Nevertheless, DIT++ lacks the detailed distinctions made when a user interacts with a search system while satisfying their information need, for example the techniques used to represent documents or information units.

\subsection{Interaction Space in Conversational Search}
\label{subsec:Interaction Space in Conversational Search}

Different schemas have been proposed for information-seeking dialogues based on dialogue acts (DAs)~\cite{searle1969speech} which try to capture the
role of an utterance.
In particular, schemas such as the COnversational Roles (COR)~\cite{stein1995structuring} and Query Request Feedback Answer (QRFA)~\cite{vakulenko2019qrfa} aim to provide the structure of a single dialogue contribution or move. In our study, we are interested in interactions between a user and SCS system in a more exhaustive manner:  
for example, utterances such as relevance feedback statements or physical actions (i.e., a mouse click to open a document). These are not covered by general-purpose DA models.

A more relevant conceptual framework was recently created by~\citet{azzopardi2018conceptualizing}. This framework combined the action and interaction space discussed in~\citet{radlinski2017theoretical} and~\citet{trippas2018informing}. The conceptual framework, therefore, is not restricted to DAs but provides an overview of the possible actions taken by either actor. We develop the action and interaction space while enriching the current frameworks.

\subsection{Natural Dialogue Study}
\label{subsec:Natural Dialogue Study}
A first step to conceptualising SCS is to explore how people interact or speak in the SCS task they are trying to accomplish~\citep{lai2009conversational}. In the case of a SCS system, one could investigate the reference interview techniques or record elicitation processes librarians undertake with information seekers~\citep{dervin1986neutral, belkin1987knowledge}. However, a more direct approach is to record a situation where people are acting as closely as possible to the task of interest~\citep{lai2009conversational}. A natural setting will encourage participants to converse more intuitively and thus provide insights into the language or vocabulary people use, their turn-taking behaviours, and the information flow~\citep{bunt1999dynamic, yankelovich2008using}. 

A natural dialogue study (NDS) supports an understanding of the accepted conversational patterns in human dialogue. Thus, more natural and usable conversational systems can be created by studying human dialogue~\cite{yankelovich2008using}. In other words, NDS helps to explore the behavioural patterns and provides insights to improve the design of the system while creating a conceptual understanding of human dialogue behaviour~\cite{bunt1999dynamic}. 

NDS is not a Wizard of Oz (WOZ) technique. In a WOZ setting, a human acts as a system while the user thinks they are interacting with a live system~\citep{gould1983composing}. 

\section{Methodology}
\label{sec:methodology}

We conducted a laboratory study to collect utterances and search interactions to develop the SCSdata. This dataset captures the utterances of two participants or actors communicating to fulfil an information need. \change{In particular, the purpose of the SCSdata is to understand how users communicate in an audio-only search setting where no screens are available to exchange information and focuses on the issues one could encounter when using such a search system.} Thus, observing how people search in this setting provides initial insight into the interactions taken~\citep{trippas2018informing}. 
To this end we conducted a study to collect a set of utterances and search interactions from two actors communicating to fulfil an information need: \textit{SCSdata}~\citep{trippas2018informing}. We developed an annotation scheme for SCS, the \textit{SCoSAS}, analysed this and validated it with inter-rater reliability; further tested it with an independent data set, Microsoft Information-Seeking Conversation data (MISC)~\citep{thomas2017MISC, thomas2018style, trippas2019data}. 
Our analysis provides insight into the interaction space and design recommendations for further research into SCS.

\subsection{Approach}
\label{subsec:Approach}

The development of spoken language datasets is a work-intensive and time-consuming process. 
Nevertheless, these datasets are invaluable for conversational modelling, as a resource for system development, or defining of vocabulary coverage~\cite{gibbon1997handbook}.
The development and evaluation of SDS is a well-studied problem and has shown that iterative analysis and assessment is needed. 

To enhance our understanding of SCS, we adopt NDS as a well-established technique used in SDS to develop a spoken language dataset and utilise qualitative analysis to identify meaningful patterns in our dataset~\cite{gibbon1997handbook, braun2013successful}. The purpose of our experimental setup is to specify the interaction possibilities in SCS. By outlining these different interactions, we provide the first step towards uncovering the details of the SCS process~\cite{gibbon1997handbook}.

\begin{figure}[htbp]
\centering
\includegraphics[trim={0 0.5cm 0 0cm},clip, width=.85\textwidth]{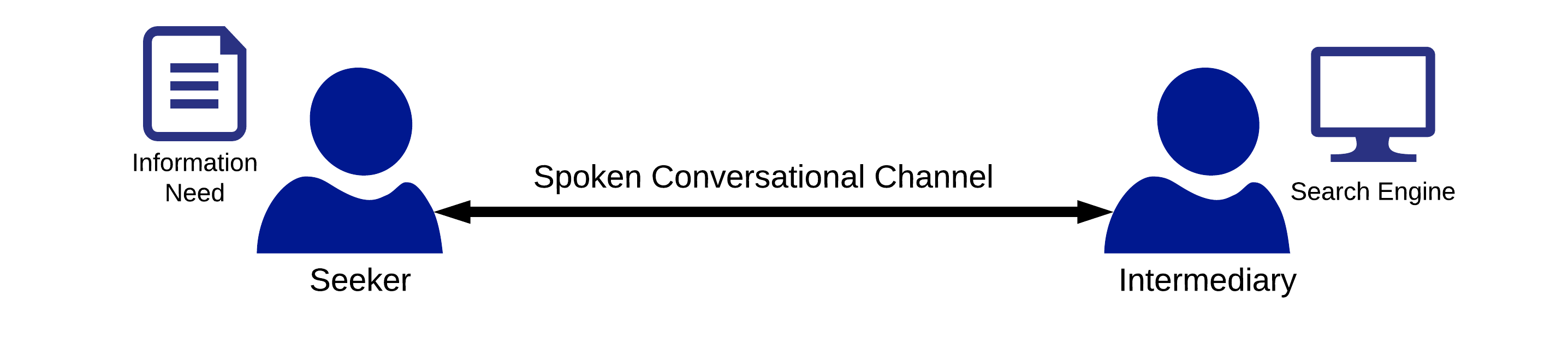}
\caption{Experimental setup.}
\label{fig:experimental_setup}
\end{figure}

Our observational study consisted of a number of \change{sessions} with two participants, where one participant acted as the \textit{Seeker} and the other participant as the \textit{Intermediary} as illustrated in Figure~\ref{fig:experimental_setup}.

The Seeker received a \textit{backstory}: a short information need statement, to motivate and contextualise the search need.\footnote{Information needs and backstories used in our experiments are listed in~\ref{sec:appendixB}.} The Seeker had to read the backstory and verbalise the information need without reading out the backstory to the Intermediary. Instead, the Seeker had to personally formulate their information need problem to convey it to the Intermediary.
The Intermediary had access to a search engine through a desktop computer. In effect, the Seeker acted as the searcher and the Intermediary simulated the audio-only interface and search system.
Participants could not access each others' tasks or search engine, were not able to see each others' facial expressions, and could only verbally communicate.
All backstories were randomised and the participant roles were randomly assigned. 

Participants completed pre-test, pre-task, post-task, and exit questionnaires, as well as a semi-structured interview. Sessions took around 90~minutes.

\subsubsection{Thematic Analysis}
\label{subsubsec:Thematic Analysis}
Thematic analysis involves identifying, analysing, and reporting patterns (themes) within  qualitative data~\citep{braun2013successful}. 
This method allows analysing qualitative data in an accessible and theoretically flexible manner, and it is often seen as a fundamental way of examining this kind of data~\citep{braun2013successful}. 
We adopted the six-step process as outlined by Braun and Clarke~\citep{braun2013successful}: 
(Step~1) familiarising self with data, 
(Step~2) generating initial codes, 
(Step~3) searching for themes, 
(Step~4) reviewing themes, 
(Step~5) defining and naming themes, and 
(Step~6) producing the report. 
\change{All steps were completed by the first author with continuous systematic feedback sessions with two other authors from Step~2 onward and a second independent annotator as validation of the full schema.}

We illustrate the two-tiered coding process in the following example from Participant~8 (Seeker). The example utterance was coded as ``Intent clarification'' as it describes the Seeker further explaining their query intent. This code is then grouped with similar codes into the ``Information Request'' sub-theme in the Task Level theme. We describe all themes and sub-themes in Section~\ref{sec:Results}.

 \begin{description}[style=multiline, labelwidth=\widthof{Intermediary long : }, font=\normalfont\textsc , leftmargin=\labelwidth, align=right]
     \item[P8 -Seeker:] Yeah, so I just want to know where it comes from \\ \textit{\small [Intent clarification]}
\end{description}

\change{To the best of our knowledge, we are the first to use thematic analysis to create of an annotation schema for SCS.}

\subsubsection{Validation of the SCoSAS schema}
\label{subsubsec:Validation of SCoSAS}

To reduce the possibility of missing important data points, we validate our coding schema in two ways. We computed (1) inter-rater reliability and code overlap, and (2) overlap and coverage based on the coding of a different dataset, the MISC\footnote{The MISC data was accessed at \url{http://aka.ms/MISCv1}.}, with our predefined codes~\cite{thomas2017MISC}.

A second independent annotator, who is familiar with information seeking and information retrieval research, recoded all utterances in the SCSdata to obtain the inter-rater reliability with Cohen's Kappa and code overlap~\cite{landis1977measurement}.
\change{The second annotator used the codebook for closed coding (i.e., the categories were already determined).}

Identifying useful actions for SCS which have not been covered in the SCoSAS provides an understanding of the scope of our coding schema. 
Therefore we apply the SCoSAS to a second and similar dataset, the MISC, to calculate the overlap and coverage~\citep{thomas2017MISC}. We took a random sample from the MISC and coded the utterances according to our dataset. 
Nevertheless, it may not be possible to achieve complete coverage with our annotations given the complexity and unexplored interactivity of a SCS information seeking dialogue~\cite{stent2000rhetorical}. In addition, achieving full coverage is difficult and often not possible to achieve~\cite{stent2000annotating}. Hence, declarations which were not covered in SCSdata received new codes according to the steps of thematic analysis.

\subsection{Data Collection Setup}
\label{subsec:Data Collection Setup}
This section introduces the experimental setup by describing the tasks used in the experiment, an overview of the participants, and the annotation steps.

\subsubsection{Task Design}
\label{subsubsec:Task Design}

We used nine search tasks and backstories from~\citet*{bailey2015user} (\ref{sec:appendixB}). These tasks covered three levels of the Taxonomy of Learning~\citep{anderson2001taxonomy}: \textit{Remember}, \textit{Understand}, and \textit{Analyse}.

\subsubsection{Participants}
\label{subsubsec:Participants}

The study involved 26 participants recruited through a mailing list.\footnote{
The protocol was reviewed and approved by RMIT University's
Ethics Board (ASEHAPP 08-16).
The mailing list is created and maintained by the Behavioural Business Lab at RMIT University: 
\url{https://orsee.bf.rmit.edu.au/public/index.php}.} 
Fifteen participants were female and 11 were male with a mean age of 30 years ($SD$=11, range 18--54).
Twenty-two participants reported being a native English speaker, and four participants said they had a high level of English proficiency. The highest level of degree held was a Master's degree. Eighteen participants reported that they were awarded a Bachelor's degree or higher and eight participants said their highest level of degree awarded was High School graduation.
The majority of participants were students (73\%), 19\% was employed, and 7\% were unemployed.
The most common fields of education were Science and Engineering (both 19\% respectively) and Law (11\%). 
Participants reported that they had been using a computer for more than ten years (85\%) and 15\% reported using a computer for 5--10 years. All participants said that they used search engines daily with the majority of participants reporting that they used a search engine more than eight times per day (54\%).

Participants rated their search skills on a 5-point scale, where 1=novice and 5=expert. Participants' mean search skills were 3.9 ($SD$=0.5), with a minimum score of 3 and a maximum of 5.

Participants' search self-efficacy was measured with the Search Self-Efficacy scale~\citep{brennan2016factor}, which contains 14 items describing different search activities. Participants indicated their confidence in completing each activity using a 10-point scale, where 1=totally unconfident and 10=totally confident. Participants' average Search Self-Efficacy was 7.3 ($SD$=1.51 and Cronbach's alpha=0.93).

Participants reported their usage of intelligent personal assistants, such as Google Now, Apple's Siri, Amazon Alexa or Microsoft Cortana. 
Four participants had never used an intelligent assistant and eight had used one a couple of times but did not use them anymore. The majority (54\%) of the participants said they used an assistant, consisting of five participants using one at least once a month and nine participants using one at least weekly.

\subsection{Data Analysis and Annotation Schema Creation}
\label{subsec:Annotation Schema Creation}

\subsubsection{SCS Dataset}
\label{subsubsec:SCS Dataset}
The SCSdata consists of 1044 turns between the 13 pairs of actors. Seekers took a total of 528 turns and Intermediaries 516. (Seekers instigated and could conclude the search, so they took 12~turns more than Intermediaries.) We recorded an average of 80 turns per pair and 26.76 turns per task.
\change{Participants exchanged 15.82 words per utterance on average with a minimum of one word per turn and a maximum of 359 words per turn. This maximum involved an Intermediary reading out a document, an action which is unusual for the dataset where the median words per turn was 9.}
\change{The SCSdata which was manually transcribed and subjected to the three-pass-per-tape policy~\citep{mclellan2003beyond}. An editor then proofread the SCSdata transcription~\citep{trippas2017protocol}.}
\change{To mitigate automatic speech recognition (ASR) problems such as out-of-vocabulary utterances, we transcribed the SCSdata manually allowing us to conceptualise the user-system interactions. For future systems, investigation will be necessary to understand the impact of ASR transcriptions on the user-system interactions.}

\subsubsection{Coding Transcriptions With Thematic Analysis to Develop SCoSAS}
\label{subsubsec:Coding of Transcriptions}

\begin{figure*}
  \centering
  \begin{tabular}{@{}c@{}}
    \includegraphics[trim={0 0.5cm 0 0.5cm}, clip, width=.95\linewidth]{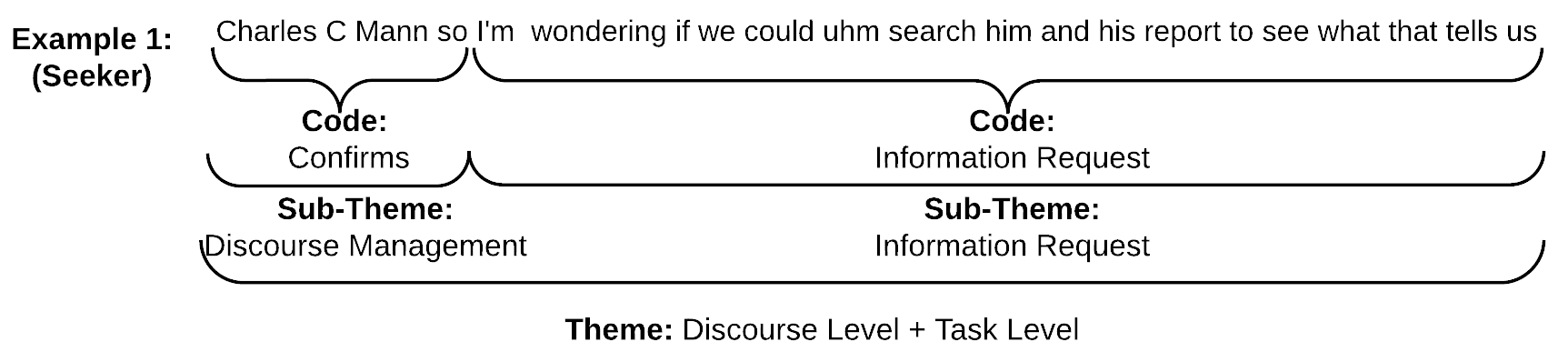} \\[\abovecaptionskip]
  \end{tabular}
  \vspace*{-0.7cm}
\vspace{\floatsep}

 \begin{tabular}{@{}c@{}}
    \includegraphics[trim={0 1cm 0 1cm}, clip, width=.95\linewidth]{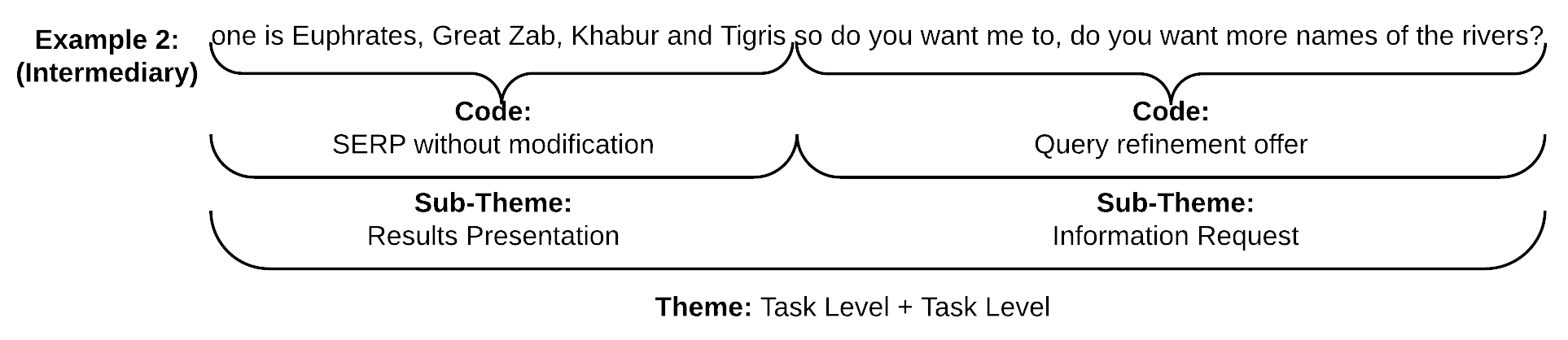} \\[\abovecaptionskip]
 \end{tabular}
  \caption{Example of coding utterances.}
  \label{fig:utterance_coding_example}
\end{figure*}

\change{We coded (i.e., labelled) our transcriptions using thematic analysis as described previously in Section~\ref{subsubsec:Thematic Analysis}. The labels of the SCSdata form the annotation schema, SCoSAS.}
We recorded both participants, and the Intermediary's screen. The recordings were synchronised and merged for transcription. 
We adopted the following steps:

\begin{description}[noitemsep]
  \item [Step 1:] Identifying when each participant spoke, i.e., identifying turns. 
  We used the approach of \textit{taking the initiative equals taking the turn}, as described by~\citet{hagen1999approach}. This means that one turn can consist of multiple moves, actions, or communication goals~\citep{tracy1990multiple}.
  \item [Step 2:] Transcribing each turn of the full dataset. 
  However, we deliberately did not eliminate any errors, false starts, or confirmations since these occur in real case voice search scenarios and to preserve the morphological naturalness of the transcription and the naturalness of the transcription structure~\cite{mclellan2003beyond, trippas2017protocol}.
  Instances where either Seeker or Intermediary was unintelligible were not transcribed but were coded as \textit{[inaudible segment]}. We assumed that if the audio recording was not clear, it was probably not clear to the other participant either. 
  
  \item [Step 3:] Designing and assigning codes to each turn with ELAN~\cite{trippas2017protocol, lausberg2009coding}. Observational notes were added. The full dataset was coded with each utterance receiving equal attention. We classified concepts from the recordings and devised a coding scheme according to the similarities across different actors. \change{The codes were designed to identify the action(s) of that particular turn, describing features of the data and defining the function of the turn (i.e., one turn/utterance could consist of multiple codes).} Thus, turns were annotated with the actions taking place. Consequently, meaningful labels were developed from the original annotations. 
\textit{Controlled Vocabulary} was added to a \textit{dictionary} which was created during coding. This dictionary was then developed into a \textit{codebook}.

  \item [Step 4:] Combining codes to themes for further analysis. Themes may consist of \textit{sub-themes} which capture specific concepts as illustrated in Figure~\ref{fig:utterance_coding_example}.
   \item [Step 5:] Checking quality assurance. Transcriptions and codes were exported from ELAN to a text file. Spelling and codes were checked.
  \item [Step 6:] Importing files into R and aggregating codes to check whether codes within a theme conceptually belonged to that theme.
  \item [Note:] 
  Steps 3--6 were conducted iteratively. This process reduced the initial 100 codes to 84 through the identification of overlapping codes.
To preserve the nuanced action described in the codes for future information seeking research, distinctions between closely defined codes were retained.
For example, the codes ``Information request'' or ``Information request within document'' were retained to identify in which section of the interaction particular information was requested. 
  \newline
  Steps 3--4 were conducted iteratively by Trippas with feedback sessions with two other authors. Random samples were investigated and compared against the coding schema, and feedback was incorporated in the next coding iteration.
  
\end{description}

\section{Results}
\label{sec:Results}

In this section, we present the themes derived from the thematic analysis together with the sub-themes which are based on the constructed codes/labels (Sections~\ref{subsec:Themes for SCS}--\ref{subsec:Theme 3}). These themes provide the characteristics of information seeking dialogues in a conversational setting, the actor's role, and the actor's relationship with the conversation. Then, we focus on the inter-rater reliability and code overlap calculations addressing the consistency of our coding schema (Section~\ref{subsec:Inter-rater Reliability}).

\subsection{Themes for Spoken Conversational Search}
\label{subsec:Themes for SCS}

Every utterance received one or more codes, based on the action taken in that utterance. For example, when an Intermediary read out a document to the Seeker, this utterance was coded with ``Scanning document''. However, when there were two actions present in one utterance -- such as when an Intermediary read out a document and then asked the Seeker whether that was useful for them -- then the utterance received two codes: ``Scanning document'' and ``Asking about usefulness''. Other examples of multiple actions and the coding of these actions are provided in Figure~\ref{fig:utterance_coding_example}.

To understand which actions are taken, we split all codes where more than one code was attached to an utterance --- thus creating atomic actions per utterance for a more natural grouping of these actions into themes and sub-themes.
We present the three themes and their corresponding sub-themes \change{and codes} as follows. The first theme, \textit{Task Level}, is related to search interactions and the topical investigation. 
The second theme, \textit{Discourse Level}, is associated with communicative functions between the Intermediary and Seeker for smooth collaboration. The third theme, \textit{Other}, consists of utterances that belong to neither the Task nor the Discourse levels.
Example utterances are provided for each sub-theme.
Tables of all the themes, corresponding sub-themes, participants (or actors), and codes are included in~\ref{sec:appendixA}.\change{\footnote{The data provided with this paper supplies all the transcripts and corresponding sub-themes and codes.}}

\subsection{Theme 1: The Task Level}
\label{subsec:Theme 1}
The Task Level theme covers search actions such as queries and search results presentation. In other words, this theme is related to the performed search task. The theme includes four sub-themes:
\paragraph*{\textbf{Information Request}}

This sub-theme covers utterances which are associated with topical information requests. 
It includes all utterances \change{with codes} which are related to forming, suggesting, refining, confirming, repeating, spelling, or embellishing information requests. The following example is of two information request sub-theme utterances:
\begin{description}[noitemsep,style=multiline, labelwidth=\widthof{Intermediary extra long: }, font=\normalfont\textsc , leftmargin=\labelwidth, align=right]
         \item[P13 -Seeker:] So which state in Australia consumes the most alcohol per person? \\ \textit{\small [Information request]}
         \item[P14 -Intermediary:] Again 2016 or the most recent information?\\ \textit{\small [Information request]}  
\end{description}

Information requests from Seekers could be expressed at any time, and they often asked for information \emph{from} a document itself, asked for meta-information \emph{about} a document or search engine results page (SERP), or provided clarification about their search intent.
Intermediaries were more likely to provide support in (re)forming the information request, for example by providing information request refinements, suggesting query expansions (i.e., whereby the initial query is augmented with query terms), or eliciting extra information.

\paragraph*{\textbf{Results Presentation}}
These sub-theme utterances convey the results from the search engine or documents: reading, interpreting, or providing an overview of a SERP or document. Only Intermediaries use this sub-theme, and the majority of Intermediary actions are linked to this sub-theme. 
In the next example the Intermediary reads out the results exactly as they were displayed in a document:
\begin{description}[style=multiline, labelwidth=\widthof{Intermediary extra long: }, font=\normalfont\textsc, leftmargin=\labelwidth, align=right]
         \item[P6 -Intermediary:] The history of valuable cinnamon. The first mention of cinnamon is in Chinese documents dating from 2800 BC. The ancient Egyptians logged cinnamon as a spice used in the embalming process...        \\ \textit{\small [Results presentation]}           
\end{description}

Other categories of utterances where Intermediaries conveyed the documents or search engine results but modified them (i.e., interpreting the results so that they would be most beneficial for the user) are also sorted in this theme.
Intermediaries modified SERPs or documents via synthesis; interpretation; paraphrase; summarisation; clarification; and comparison.

\paragraph*{\textbf{Search Assistance}}
This sub-theme captures interactions where the Intermediary assisted the search process by providing explicit search suggestions, advice,  or relevance judgements:
\begin{description}[style=multiline, labelwidth=\widthof{Intermediary extra long: }, font=\normalfont\textsc , leftmargin=\labelwidth, align=right]
         \item[P2 -Intermediary:] there is a lot on health benefits conversation uhm [long pause] I don't see how some of these are relevant     \\ \change{\textit{\small [Results presentation + search assistance]}}         
\end{description}

In contrast to directly providing assistance, Intermediaries also asked how to help the Seekers in their search process. This was seen by asking about the usefulness of a result, requesting spelling, or suggesting a different search engine.

Additionally, this sub-theme captures the Seeker explicitly asking for assistance during their search session: for example by asking for recommendations or judgements on whether they covered enough of the information space.

\paragraph*{\textbf{Search Progression}}
This sub-theme is only used by the Seeker to provide feedback on progress: for example by giving performance feedback, rejecting search results, or informing whether they found enough information for a topic:
\begin{description}[style=multiline, labelwidth=\widthof{Intermediary extra long: }, font=\normalfont\textsc , leftmargin=\labelwidth, align=right]
         \item[P15 -Seeker:] OK that's probably enough information \\ \textit{\small [Search progression]}
\end{description}

In summary of this first theme, the Seeker evoked all sub-themes in the Task Level theme except the Results Presentation sub-theme. The Results Presentation sub-theme was only used by the Intermediary allowing them to present found information from the search engine to the Seeker. The Intermediary also evoked all sub-themes except the Search Progression sub-theme, which was used by the Seeker to provide feedback to the Intermediary.

\subsection{Theme 2: The Discourse Level}
\label{subsec:Theme 2}

The Discourse Level theme covers aspects which are not linked to performing a topical (search) task
but instead are concerned about the audio channel between participants. 
The Discourse Level theme consists of four sub-themes: 
\textit{Discourse Management} which allows the conversation to take place between the actors; 
\textit{Grounding} captures interactions for creating mutual knowledge, beliefs, and assumptions between the two actors~\citep{traum1999speech, clark1991grounding}; 
\textit{Navigation} which covers the communications of moving around web pages, documents, and browser tabs; 
and \textit{Visibility of System Status} which allows actors to provide feedback on what is happening throughout interactions.

\paragraph*{\textbf{Discourse Management}} 
This sub-theme includes conversational coherence and cohesion between the actors~\citep{schiffrin1985conversational}. In other words, the utterances in this sub-theme are part of the communication between the actors to check whether the other actor has understood a message. In our dataset, these discourse building utterances are independent of the participant role. For example, both Seeker and Intermediary confirmed, checked, or asked to repeat and repeated utterances as illustrated in the snippet below.
\begin{description}[noitemsep, style=multiline, labelwidth=\widthof{Intermediary extra long: }, font=\normalfont\textsc , leftmargin=\labelwidth, align=right]
        \item[P1 -Seeker:] So uhm can you go and change the search question to effectiveness of uhm... passenger and baggage screenings at airport 
         \item[P2 -Intermediary:] Passenger and      \\ \textit{\small [Discourse management]}  
         \item[P1 -Seeker:] Baggage  \\ \textit{\small [Discourse management]}
\end{description}

Often an information request was echoed or either actor confirmed a command. These discourse actions are crucial to have a meaningful conversation, for example indicating that one actor has understood the other.

\paragraph*{\textbf{Grounding}}

\textit{Grounding in communication} as described by~\citeauthor{clark1991grounding} is ``sharing and synchronising mutual beliefs and assumptions'' and is fundamental for communication between actors~\cite{clark1991grounding}. The two actors' mental model of each others' beliefs needs to be continuously updated to coordinate the build of a mutual understanding.

We observed utterances belonging to this particular sub-theme which was used by Seekers to coordinate the shared information or common ground~\cite{clark1991grounding}. Seekers summarised or paraphrased the information given to them and created a bigger picture of the search results as a way of synchronisation. 

Through this dynamic process, Seekers provided insight on what they understood from the information provided; Intermediaries then knew whether the information was correctly conveyed.

\begin{description}[style=multiline, labelwidth=\widthof{Intermediary extra long: }, font=\normalfont\textsc , leftmargin=\labelwidth, align=right]
         \item[P14 -Intermediary:] [...] yeah 20 to 29 is the most high risk drinking people in Australia for alcohol related harm... I don't know what that means about consumption
         \item[P13 -Seeker:] Yeah so they consume a lot \\ \textit{\small [Grounding]}
\end{description}
	
Grounding differs from Search Progression and Discourse Management. While Grounding involves sharing the beliefs and values of the information, Search Progression is concerned with the feedback on the search task progress and Discourse Management is related to effective information transfer.

The Grounding sub-theme was only seen in Seekers' utterances. This is because Intermediaries, by having the information to hand, summarised results presented and they did not need to confirm or share their beliefs or meaning of the content. As such, their utterances are captured by Results Presentation.

\paragraph*{\textbf{Navigation}}
Navigational utterances allow \change{actors to progress} the task by manoeuvring around the online information space. 
We observed Seekers navigate the search results by instructing the Intermediaries. Seekers asked to access specific sources, navigated between documents, singled out particular documents, and read more from a document or the next document:
\begin{description}[style=multiline, labelwidth=\widthof{Intermediary extra long: }, font=\normalfont\textsc , leftmargin=\labelwidth, align=right]
         \item[P9 -Seeker:] Uhm maybe uhm can you go into the result [...] that mentions how uhm outsourcing damages the industry \\ \textit{\small [Navigation]}
\end{description}

\paragraph*{\textbf{Visibility of System Status}}

Seekers asked the Intermediaries to provide information on what was occurring throughout the interactions: for example, whether what they asked for was fulfilled, or what the results were. 
Intermediaries provided feedback on what was taking place on their side of the conversation by giving input on what was happening (i.e., keeping each other informed~\cite{nielsen2005ten}) if they had seen certain items before, or by way-finding (i.e., orienting where they were positioned). For example:

\begin{description}[noitemsep, style=multiline, labelwidth=\widthof{Intermediary extra long: }, font=\normalfont\textsc, leftmargin=\labelwidth, align=right]
         \item[P25 -Seeker:]  Oh TIBER sorry Tiber yeah\\ \textit{\small [Discourse management]}
         \item[P26 -Intermediary:] Yeah uhm just searching just one second     \\ \textit{\small [Visibility of system status]} 
         \item[P25 -Seeker:]  Any luck?\\ \textit{\small [Visibility of system status]}
\end{description}

\subsection{Theme 3: Other}
\label{subsec:Theme 3}

Five utterances from the Seeker were not classified in any of the above \mbox{(sub-)themes}. Two of these utterances were disfluencies from the Seeker, one utterance was where the Seeker provided information about the search engine, one utterance was asking if the Seeker was allowed to embellish a query, and the last unclassified utterance involved the Seeker offering to spell a word. These five categories were not classified after much deliberation and given the theme ``Other'' instead.

\hspace{\parskip} 

To finalise the themes examination, an overview of the themes and sub-themes used by each actor is presented in Table~\ref{tab:Themes and sub-themes}. The development of the classifications in themes, sub-themes, and codes form the basis of the Spoken Conversational Search Annotation Schema (SCoSAS).
{\renewcommand{\arraystretch}{.8}
\begin{table}[htp]
\centering
\smaller{}
\caption{Themes and sub-themes used by different actors}
\label{tab:Themes and sub-themes}
\begin{tabular}{llcc}
\toprule
\textbf{Theme} & \textbf{Sub-theme} & \multicolumn{1}{l}{\textbf{Seeker}} & \multicolumn{1}{l}{\textbf{Intermediary}} \\ \midrule
\multirow{4}{*}{Task Level}           & Information Request           &     \checkmark      &     \checkmark                                  \\
     & Results Presentation        &                                   &    \checkmark      \\
    & Search Assistance        &    \checkmark              &   \checkmark       \\
    & Search Progression \ignore{or Meta-discussion}       &    \checkmark                  &          \\\midrule
\multirow{4}{*}{Discourse Level}  & Discourse Management         & \checkmark     & \checkmark                                     \\ 
 & Grounding & \checkmark &       \\
 & Navigation & \checkmark &       \\
 & Visibility of System Status & \checkmark & \checkmark      \\ \midrule
Other &  & \checkmark &       \\ 
\bottomrule
\end{tabular}
\end{table}
}

\subsection{Inter-rater Reliability and Code Overlap}
\label{subsec:Inter-rater Reliability}

\change{As part of the validation of the SCoSAS, we calculate the inter-rater reliability and code overlap (i.e., the utterance code overlap and code usage overlap).}

The first author (Assessor~1) created the codes as described above. A second independent researcher (Assessor~2) used the codebook for closed coding of all utterances in the SCSdata. \change{The inter-rater reliability on code level (i.e., atomic action identified on an utterance) was moderate (Cohen's $\kappa=0.59$)~\citep*{landis1977measurement}, and substantial at the sub-theme level (i.e., classification based on the code) ($\kappa=0.71$).}

The overlap of codes was high with 90\% of the predefined codes being used by both assessors. More precisely, Assessor~1 applied 84 different codes consisting of 41 codes for the Seeker and 43 for the Intermediary. Assessor~2 used 76 codes, 38 codes for the Seeker and 38 for the Intermediary as seen in Table~\ref{tab:Independent Assessors' Code Overlap}.

{\renewcommand{\arraystretch}{.8}
\begin{table}[htp]
\centering
\smaller
\caption{Independent Assessors' Code Overlap}
\label{tab:Independent Assessors' Code Overlap}
\begin{tabular}{lcc}
\toprule
                                                                                 & \textbf{Assessor 1} & \textbf{Assessor 2} \\ \midrule
Total number of utterances                                                       & 1,044       & 1,044       \\
Total number of codes used                                                     & 84        & 76        \\
Total number of codes for Seeker                                            & 41               &  38         \\
Total number of codes for Intermediary                                  & 43               &  38         \\
Unused codes                                                                           & 0          & 8 (10\%)        \\ 
 \bottomrule
\end{tabular}
\end{table}}

The eight codes used by Assessor~1 but not Assessor~2 could potentially be consolidated in a future refinement.

\section{Validation of SCoSAS} 
\label{sec:Validation of SCoSAS}

\change{To explore the extent to which SCoSAS covers SCS interactions, we applied the SCoSAS to a subset of interactions from a second dataset, the MISC~\citep*{thomas2017MISC}.}

\subsection{Using the MISC dataset to validate SCoSAS}
\label{subsec:Validation of SCoSAS with MISC}

As with the SCSdata, the MISC dataset~\citep{thomas2017MISC} is a collection of recorded information-seeking conversations between a Seeker and an Intermediary. MISC contains audio and video recordings with \change{automatic speech recognition transcriptions of these recordings.} \change{We coded the MISC dataset according to our predefined codes to investigate which actions were covered or not covered by our coding schema.} Thus, by using our predefined codes, we validate the coverage (i.e., is there an action applicable for every situation) and overlap (i.e., is there a situation where more than one action could be relevant).

We selected a random set of four participant pairs for the labelling: participants 1--2, 7--8, 19--20, and 27--28. The MISC setup has five tasks for each pair, of which one is a practice. We labelled the four remaining tasks per participant pair for a total of 16 task-instances. \change{The labelling was completed by Trippas.}

\change{Among} the four pairs, we have a total of 701 turns with an average of 175.25 turns per pair and an average of 43.81 turns per task. However, 5\% of the total turns in the MISC transcriptions were inserted by the ASR and were not present in the audio. These turns were ignored which means that a total of 666 turns were labelled on code-level with an average of 166.5 turns per pair and an average of 41.62 turns per task.

\subsection{Differences Between the SCSdata and MISC}
\label{subsec:Differences Between the SCS and MISC Datasets}
The setup and instructions between the SCSdata and MISC protocols were marginally different, which led to differences in the data. We provide an overview of the differences in this section; a fuller account is in~\citet{trippas2019data}.

\paragraph{Setup of SCSdata and MISC}
\label{subsubsec:Setup of MISC and SCS}

As with SCSdata, MISC Seekers did not have access to any information source, but received an information need which they relayed to an Intermediary over an audio connection. Unlike SCSdata, MISC Seekers were allowed to read out the need as given, but were also asked to record an answer.

\paragraph{Transcription Differences}
\label{subsubsec:Transcription Differences}
The MISC dataset was transcribed using ASR in contrast to the SCSdata which was manually transcribed, subjected to the three-pass-per-tape policy, and proofread by a professional editor.

\change{The ASR was prone to error, in particular ``recognising'' utterances such as ``thank you'' that were not in the audio. The following snippet of a conversation is an illustration: speakers appear more polite than they were.}
        
        \begin{description}[noitemsep, style=multiline, labelwidth=\widthof{Intermediary extra long: }, font=\normalfont\textsc, leftmargin=\labelwidth, align=right, itemsep=0mm]
                    \item[P20 -Intermediary:] [...] She wanted them to donate to charity
                    \item[P19 -Seeker:] Thanks \\ \textit{\small [Utterance not present in audio]}
                    \item[P20 -Intermediary:] To provide clean water // and she um                  
                    \item[P19 -Seeker:] Thank you \\ \textit{\small [Utterance not present in audio]}
        \end{description}
        
We encountered an occasion where the researcher interfered due to a technical issue and sections where the ASR created many unnecessary turns between the actors because it falsely believed that someone was talking. We excluded these utterances from this analysis.

\subsection{Creating Comparable Datasets} 
\label{subsec:Creating Comparable Datasets}
\label{subsec:Utterance Labelling}

The MISC dataset contains ASR errors and the subset we used did not include screen capture video. We labelled MISC at the code level. However, subtleties such as whether an Intermediary was reading from a SERP or document could not be distinguished without screen captures and all Results Presentation utterances were labelled just with that sub-theme.

\subsection{Results: Overlap and Coverage Between SCSdata and MISC} 
\label{subsec:Overlap and Coverage Between SCS and MISC Data}

We are interested in the number of actions shared between the SCSdata and MISC, and where actions are different.
After collapsing all Results Presentation utterances to the sub-theme level, for compatibility with MISC, SCSdata used 66~distinct codes: 41 from Seekers and 25 from Intermediaries. MISC used 31 for Seekers and 18 for Intermediaries (Table~\ref{tab:SCS and MISC Dataset Descriptives}).

{\renewcommand{\arraystretch}{.8}
	\begin{table}[ht]
		\centering
		\smaller
		\begin{threeparttable}
			\caption{SCSdata and MISC Descriptives}
			\label{tab:SCS and MISC Dataset Descriptives}
			\begin{tabular}{lcc}
				\toprule
				& \textbf{SCSdata} & \textbf{MISC subset} \\ \midrule
				Total number of utterances                                                       & 1044       & 666       \\
				Total number of unique codes*                                                  & 66*        & 49*        \\
				Unique codes Seeker                                                              &   41           &    31     \\
				Unique codes Intermediary                                                 &       25       &      18   \\
				\bottomrule
			\end{tabular}
			\begin{tablenotes}
				\smaller
				\item *NOTE: Due to insufficient details, utterances which were related to presenting results were aggregated to the Results Presentation sub-theme level. The SCSdata's unique number of codes without aggregation of the Results Presentation is 84.
			\end{tablenotes}
		\end{threeparttable}
	\end{table}
}

To label MISC, we needed 49 codes, of which 35~were used in both sets; 14~additional codes were needed to cover actions not seen in SCSdata. These additional codes however were infrequently used and 94\% of MISC utterances could be coded with SCoSAS. The 14~additional codes, covering 6\% of utterances, are summarised in Table~\ref{tab:MISC codes which are not present in SCS}, and we discuss these below.

{\renewcommand{\arraystretch}{.8}
\begin{table}[ht]
\centering
\smaller
\caption{Set difference between SCSdata and MISC.}
\label{tab:MISC codes which are not present in SCS}
\begin{tabular}{lcc}
\toprule
   \textbf{Code}               & \textbf{Actor} & \textbf{Nr used} \\ \midrule
Chitchat                                                      &   \multirow{8}{*}{Seeker}      &   1  \\
Communication about the task                    &       &   2    \\
Decision offloading                                        &        &   1        \\ 
Feedback on writing down the answer for the given task                                      &        &  3       \\
Negotiation                                                  &        &  7      \\
Rejects spelling offer                                  &        &   1       \\
Requests spelling                                          &        &   1        \\
Uncertainty expression of  what to search                  &        &   2      \\
\midrule
Chitchat                                                    & \multirow{6}{*}{Intermediary}       &  5          \\
Enough information?                               &        &  9          \\
Negotiation                                                    &        &  6         \\
Offers to spell                                             &        &    1        \\
Spells                                                    &        &  5        \\
Too many results to sum up               &        &    1       \\ \midrule
\begin{tabular}[c]{@{}l@{}}Total number of instances of code \\ used by MISC and not by SCSdata\end{tabular}      & & 45 (6\%)\\
 \bottomrule
\end{tabular}
\end{table}
}

\paragraph{\textbf{Chitchat or Negotiation}} 
\label{subsubsec:Chitchat or Negotiation}
We encountered new types of utterances in the MISC where the actors were negotiating or chitchatting. The negotiation utterances were used to bridge differences and reach agreements~\citep*{zuckerman2015first}. Examples include instances where actors share their own experiences about particular topics or subjects. 
However, this is not to be confused with the already defined Grounding sub-theme which covers utterances from the Seeker expressing their beliefs and values of information provided by the Intermediary.

Chitchat and negotiation utterances have greater overlap between speakers, meaning that more than one actor at a time is speaking~\cite{schegloff2000overlapping}. For example, the following utterances overlapped while the Seekers and Intermediary negotiated their shared understanding of non-traditional medicine:
\begin{description}[noitemsep, style=multiline, labelwidth=\widthof{Intermediary extra long}, font=\normalfont\textsc, leftmargin=\labelwidth, align=right, itemsep=0mm]
          \item[P1 -Seeker:] I think herb sounds more like // not \\ \textit{\small [Negotiation]}
          \item[P2 -Intermediary:] More like medicine \\ \textit{\small [Negotiation]}              
           \item[P1 -Seeker:] I think it sounds more like naturopathic but that fits it \\ \textit{\small [Negotiation]}
\end{description}

Participants seemed forthcoming in sharing their own opinions and experiences. The following example is from an Intermediary who shares her own travel experiences which are related to the task:
\begin{description}[noitemsep, style=multiline, labelwidth=\widthof{Intermediary extra long}, font=\normalfont\textsc , leftmargin=\labelwidth, align=right]
        \item[P8 -Intermediary:] That's what I love to do actually when I traveled all the public transportation and all sorts of continents \\ \textit{\small [Chitchat]}
\end{description}

\paragraph{\textbf{Communication about the task}} 
\label{subsubsec:Communication about the task}

SCSdata participants were instructed not to share the given search task but instead rephrase request. However, for MISC, participants were allowed to read out their search task. This led to Seekers talking informally about the task itself. For example,

 \begin{description}[style=multiline, labelwidth=\widthof{Intermediary long : }, font=\normalfont\textsc , leftmargin=\labelwidth, align=right]
     \item[P1 -Seeker:] Yeah the task is a bit // um very generalised so um
\end{description}

\paragraph{\textbf{Agency and Decision Offloading or Taking Control}} 
\label{subsubsec:Agency and Decision Offloading or Taking Control}
In MISC, both Seeker and Intermediary share the same information and underlying ideas of what they need to search for. This created an equal level of collaboration between the two actors. However, it also allowed the Intermediary to instantiate more agency. \change{In contrast, Intermediaries in the SCSdata acted more as the interface between the Seeker and the found information.}

We noticed this idea of agency throughout the subset of the MISC in actions resulting in the following codes ``Enough information?'' (Intermediary), ``Too many results to sum up'' (Intermediary), and ``Decision offloading'' (Seeker). For example, the Intermediaries suggested that a search task has been finished \textit{``excellent, so we are finished...''} (P8) or that they are not going to sum up all the results. The Seekers also handed over the decision making to Intermediaries: e.g. \textit{''it's up to you [ed.\ if we look at the other site or not]''} (P20).

\change{\citet{trippas2018informing} suggested decision offloading and taking control may be an artefact of the linear audio channel. The system thus creates a cost-benefit estimation of whether further information from the Seeker is required and therefore receives more autonomy. Simultaneously the Seeker can suggest the system to make the discussion due to the limited knowledge which can be transferred in the audio channel.}

\paragraph{\textbf{Writing Down the Answer}} 
\label{subsubsec:Feedback on Writing Down Answer for the Given Task}

Seekers in the MISC setup were asked to write down an answer to their given information seeking task. Seekers' utterances therefore include how they were progressing with the writing. We also encountered several different instances of spelling actions in MISC which we had not encountered in the SCSdata.

\paragraph{\textbf{Uncertainty in What to Search}} 
\label{subsubsec:Uncertainty expression of what to search}

As mentioned, MISC Seekers were allowed to read their search task aloud. Here, the Seeker is expressing their confusion with the task:

 \begin{description}[style=multiline, labelwidth=\widthof{Intermediary extra long: }, font=\normalfont\textsc , leftmargin=\labelwidth, align=right]
     \item[P19 -Seeker:] I am not sure what you're supposed search
\end{description}

This could be interpreted as identifying a gap in the Seeker's knowledge. However, the information need expression is not formalised~\citep{taylor1962process}. Recently, \citet{trippas2018informing} suggested that formulations of needs in SCS do not conform to the typical textual query. In a voice environment, users can use natural language to describe their search, and the information request may not go through~\citeauthor{taylor1962process}'s four stages of information need~\cite{taylor1962process}. 
In the above example, we could even argue that users may now have the freedom to tell the system that they have identified a gap in their knowledge before yet formalising their problem.

\subsection{Discussion of SCoSAS Validation} 
\label{subsec:Discussion of SCoSAS Validation}

The majority of the codes (71\%) seen in MISC overlapped with the SCoSAS, and the novel codes only covered 6\% of utterances. Some of the new codes were not encountered in the SCSdata due to the difference in experimental setups, such the array of possible spelling requests, suggestions, or declines. These would be valuable expansions to the SCoSAS.

\section{Discussion}
\label{sec:discussion}

In this work, we used a qualitative analysis approach to uncover the range of possible interactions of information moves for Seekers and Intermediaries in a SCS setting. We constructed insight into the conversational structure of information seeking processes. To do so, we first created a spoken dataset, the SCSdata, and then derived an annotation schema for conversational search via a thematic analysis approach. Finally, we evaluated these actions against a similar dataset as evaluation. 

\subsection{Schematic SCS Themes Model}
\label{subsec:Schematic SCS Themes Model}

\begin{figure}[t!]
\centering
\includegraphics[width=.84\textwidth,
trim=4cm 1.7cm 4.2cm 1.5cm,clip]{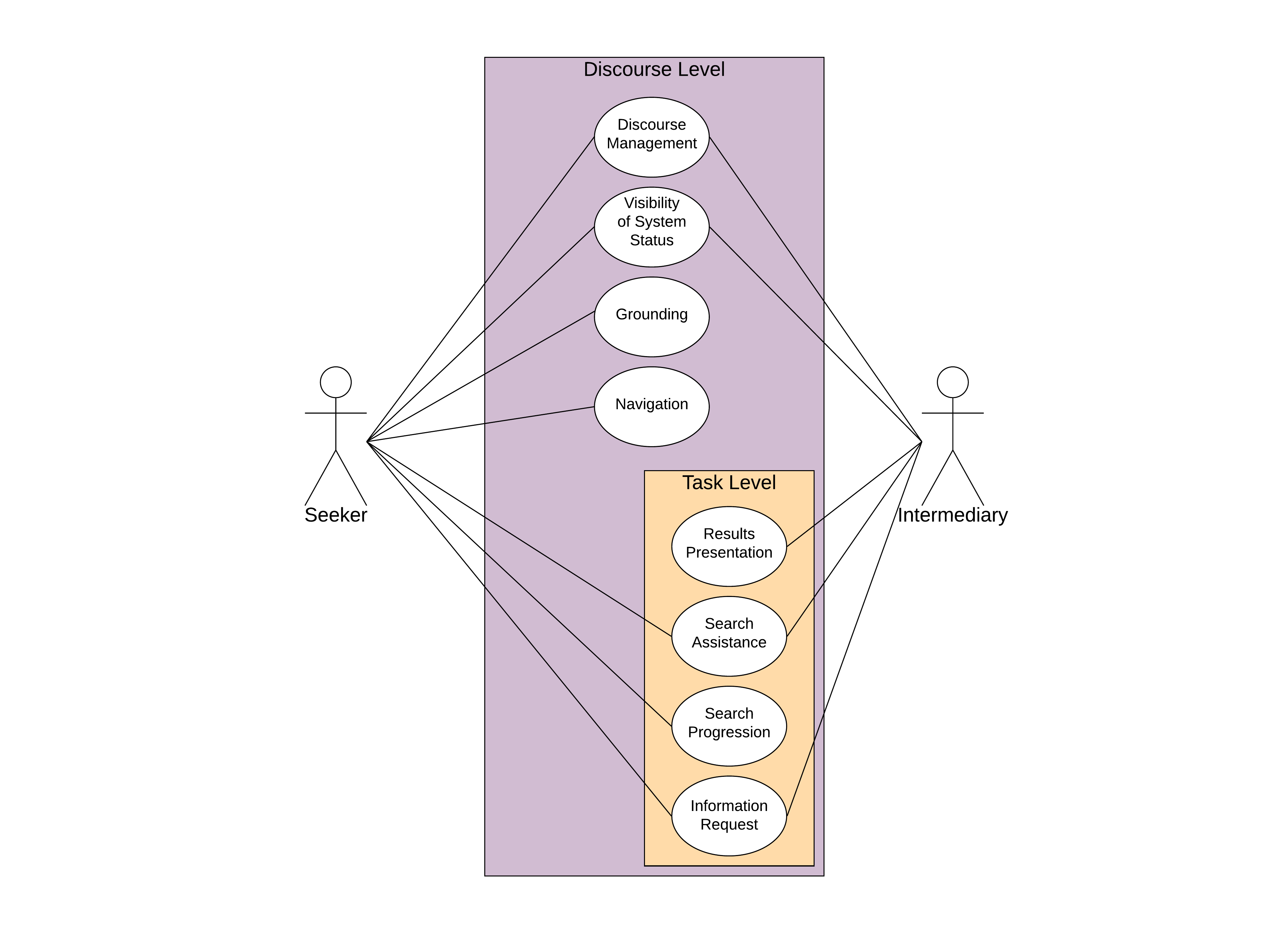}
\caption{Schematic Model of Themes and Sub-themes used by each actor.}
\label{fig:Schematic Model of Themes and Sub-themes}
\end{figure}

The SCoSAS presents the Task Level at the centre of conversations. 
The Discourse Level surrounds this, representing the statements which are about the mechanism, not the task (see Figure~\ref{fig:Schematic Model of Themes and Sub-themes}). The Discourse Level would still exist if the search task is changed to a different, unrelated, task. Previous research in communication goal studies suggests a similar two-tiered model~\citep{tracy1990multiple,bunt1999dynamic}.
\change{Furthermore, the goal studies community argues that ordinary discourse is segmented in different types of goals such as communicative functions or interaction outcomes which is similar to our two themes of Task and Discourse.
\citet{bunt1999dynamic} provided a two-tiered model where general information dialogues consist of two motivations, that is, one tier was concerned about the task communication and the second.}

Our results highlight the importance and need for integrating discourse in SCS systems and to the best of our knowledge, discourse functions are yet to be integrated in information seeking models~\cite{stein1995structuring, vakulenko2019qrfa}. Furthermore, including these discourse utterance inherently creates a system which interacts in a mixed-initiative information seeking communication (the system can ask for clarification and thus takes initiative). Such mixed-initiative dialogue is a requirement of what makes a SCS system truly conversational~\cite{culpepper2018research, trippas2018informing}.

This first attempt at creating an interaction model of two actors in a SCS setting may not have included all possible future actions. One action we believe may be observed in a real system is for the user to test the abilities of the system or access the settings of the system itself. This might be coded as a System Level theme, overlapping both Discourse and Task Levels.

\subsection{Design Recommendations for SCS Systems}
\label{subsec:Design Suggestions}

Our analysis leads to some design recommendations for SCS systems.

\paragraph{\textbf{Integrating Search Assistance}}
\label{subsubsec:Integrating Search Assistance}
Search assistance is integrated in many different ways in browser-based search, for example by query or spelling suggestions. We could extend these assistance functions to include the system providing relevance feedback to the user about a given document, suggesting to move on, or even asking about the usefulness of a given result. These pro-active features can become part of the model of the user preferences given the interaction history~\citep[c.f.][]{radlinski2017theoretical}.

\paragraph{\textbf{Grounding as Relevance Feedback}}
\label{subsubsec:Grounding as a Relevance Feedback}

\change{Grounding (i.e., discourse for the creation of mutual knowledge and beliefs) is when participants in a conversation engage in a specific discourse activity to share their mutually understood utterances~\cite{clark1991grounding}. We observed grounding actions in the SCSdata. For example, Seekers provided indirect feedback by reciting their interpretation of the found results. This grounding process could enable a future SCS system to better understand a user’s awareness of the results or information space, including helping the SCS system to disambiguate a users’ information need.}

\paragraph{\textbf{Visibility of System Status}}
\label{subsubsec:Visibility of System Status}
Visibility of system status enables greater control, explainability, and transparency of the system processes and outputs~\citep{culpepper2018research}. However, providing constant feedback on what is happening in a system is not convenient in a spoken environment and will overwhelm the user with too much unnecessary information. It will be essential to understand which aspects should be given to the user.
At any point in time, the system should be able to disclose how it retrieved or computed specific information.

\paragraph{\textbf{Navigation}}
\label{subsubsec:Navigation}

Navigational interactions often contain a mix of selecting links on a web page or using backtracking techniques such as the back button or history list~\citep{fu2007snif}. These navigation actions have been extensively studied in a text environment~\citep{catledge1995characterizing,kellar2007field}. Recently~\citet{azzopardi2018conceptualizing} and~\citet{trippas2018informing} recognised the importance of these actions in a conversational search setting. Instead of interacting with lists in a spoken environment as often is done in a SDS, users can freely navigate in a multi-dimensional information space.

\change{Navigational interactions in this study may have been influenced by how a system works through back buttons and links. In the future we expect the navigation space to expand together with the adoption and creation of conversational systems.
Being able to present a traceable history also provides further transparency for the user and supports the explainability of the system. 
For example, breadcrumbs could refer to previous information spaces or provide summaries of information the user visited instead of titles of documents as in a browser-based back-button action.}

\subsection{Evaluating Existing Search Behaviour Models with SCoSAS}
\label{subsec:Contrasting Existing Search Behaviour Models with SCoSAS}
\change{To our knowledge many well-known models such as~\citeauthor{belkin1980anomalous}'s ASK~\cite{belkin1980anomalous}
or~\citeauthor{marchionini1997information}'s ISP~\cite{marchionini1997information} do not include the system's  ``responsibility'' of interacting with the user and thus do not capture all SCS behaviours.}

\change{Other models, such as~\citeauthor{sitter1992modeling}'s COR model~\cite{sitter1992modeling}, \citeauthor{belkin1995cases}'s scripts~\cite{belkin1995cases}, or the recently proposed QRFA model by~\citet{vakulenko2019qrfa} encompass the interaction between two actors. 
However, these models either lack the flexibility of the speech aspect, such as multiple moves in one turn, or are based on broad DA categorisations.
Additionally, the broad DA categorisation only provides a high level insight of the actions users take while the SCoSAS discloses more refined details of the users' and systems' state in each turn.}

\change{Finally, \citeauthor{saracevic1997stratified}'s stratified model includes the system as an active participant in the information seeking process~\cite{saracevic1997stratified}. Furthermore, Saracevic specifies that the process consists of a dialogue between the two actors. He also mentions that  the dialogue can be used for not only ``searching'' utterances but also for a number of ``other engagements'' beyond the searching, for example, obtaining and providing different types of feedback, judgements, or states. In the SCS model, we also identify the system as an active participant throughout the search process, which is in itself a conversation.
In addition, the ``other engagements''~\citeauthor{saracevic1997stratified} mentions could be interpreted as our Discourse Level interactions, such as our identified grounding utterances.
Furthermore, the stratified model could be used to illustrate the effect of the audio-only interaction channel limitation. That is, \citeauthor{saracevic1997stratified} says that a weak point in the system could hamper the desirable outcome for the search process~\cite{saracevic1997stratified}. The stratified model and the schematic SCS themes model may be complementary for the abstraction of a SCS process.}

\subsection{Future Extensions}
\label{subsec:Limitations}

\textit{Human to human interaction: }Human--human interaction may differ from the human--machine interactions we really want to model. We plan to conduct further studies to test our hypotheses in a human--machine interaction setting. \change{Thus, further research will investigate if the mindlessness projection of users' own believes and expectations of computers transfers to SCS~\cite{nass2000machines}.}

\textit{Laboratory setting: }Participating in a laboratory setting \change{influences} the participants' behaviour. Even though this study was conducted in a laboratory setting, we believe findings will apply to a day-to-day environment. 

Investigating the information needs for SCS which arise in a natural setting will be crucial to develop natural systems. This will include understanding the different information needs and creating new taxonomies for these needs.

\textit{Taking initiative equals one turn: }Our coding schema allows for coding
per turn since we segmented the users' utterances with the idea that taking the initiative equals one turn. This means that slight subtleties inside a turn may be lost such as long pauses. However, we believe this 
was necessary to understand the broader context of SCS.

\change{This study should be interpreted within the context of its limitations. Firstly, it is possible that cross-coding a larger dataset from the MISC may have added further sub-themes. However, we feel that the discrepancies identified through the current cross-coding are due to differences in experimental set-up rather than to substantial content differences.}

\subsection{Informing Wider Research Agendas}
\label{subsec:Informing Wider Research Agendas}

Existing systems and models have difficulties with multi-turn actions, utterances which consist of multiple moves, or intent extraction. In this paper, we attempted to better understand these unique features of SCS by creating a labelling schema and schematic model of these labels. 
Our model and annotation schema are a more in-depth study than prior preliminary models~\citep{trippas2017people} or conceptual framework~\citep{azzopardi2018conceptualizing}.
While the labelling schema developed in this paper focuses on the interaction space of conversational search, we expect it will be useful for non-search related or discourse actions. 
\change{The implications of this analysis are many. Firstly, this analysis can support the feature extraction of particular utterance-types, or assist with the engineering and evaluation of conversational retrieval. The analysis can also be used for language modelling of information seeking conversations and the development of results presentation strategies.}

\section{Conclusions}
\label{sec:conclusions}

In this paper, we address the challenge of \change{spoken conversational search} (SCS), where no screens are available and user--system interactions are entirely voice-based. After identifying the limitations of existing information seeking models, we used a qualitative analysis approach to explore how people interact in an audio-only communication search setting. 
We created the first dataset for SCS (SCSdata), defined a labelling set identifying the interaction choices for this dataset (i.e., the SCoSAS annotation schema), and translated these interactions in a schematic model. This approaches both actors in the seeking process, the Seeker and the Intermediary, as equal, leveraging multi-turn activities and multi-move utterances. The validation of SCoSAS using an independent dataset demonstrated high overlap and coverage.

\change{Furthermore, our transparent annotation process contributes by strengthening the analysis and the methodological foundations of annotation schema development.}

The significance of this paper is twofold: we (i) develop a classification schema, (ii) test and validate this schema, and (iii) provide a transparent annotation schema process.
Furthermore, our contributions highlight the need of new models for SCS, especially integrating discourse. The resources described and validated in this paper -- including the SCoSAS annotation schema -- also allow us to suggest possible extensions of the schematic model and inform the design of SCS systems in the future.

\section{Acknowledgements}
The authors would like to thank the participants who took part in the study. This research was partially supported by Australian Research Council Projects LP130100563 and LP150100252, Real Thing Entertainment Pty Ltd, and JSPS KAKENHI JP19H04418.

\def\bibsection{\section*{References}}
\setlength{\bibsep}{0pt plus 0.5ex} 
\bibliographystyle{abbrvnat}
\bibliography{conversational_model}

\appendix
\clearpage
\newpage
\section{Themes, Sub-themes, and Codes in SCoSAS}
\label{sec:appendixA}

\subsection{Theme 1: Task Level}
\label{appendix:Theme 1: Task Level}

{\renewcommand{\arraystretch}{.8}
\begin{table}[ht]
\centering
\caption{Information Request Codes (Seeker)}
\label{tab:Information Request Codes (Seeker)}
\adjustbox{max width=1\textwidth}{
\begin{tabular}{lllp{6cm}c}
\toprule
\textbf{Theme} & \textbf{Sub-theme} & \textbf{Actor} & \textbf{Code}                       & \textbf{Frequency} \\ 
\midrule
\multirow{24}{*}{Task Level}   & \multirow{24}{*}{Information Request} & \multirow{12}{*}{Seeker} & Automated repetitive search\change{~\cite{trippas2018informing}}         & 3         \\
               &       &               &  Definition explanation \change{(or clarification, i.e., intent clarification)}              & 1         \\
               &       &                & \change{(Information request for a)} Definition lookup or person         & 1         \\
               &        &              & \change{(Requesting)} Information about document\change{~\cite{trippas2018informing}}          & 6         \\
               &        &              & Information about SERP overview     & 2         \\
               &        &             & Information request                 & 67         \\
               &        &              & Information request within document & 80         \\
               &        &               & Information request within SERP     & 15         \\
               &       &               & Initial information request         & 39         \\
               &       &               & Intent clarification                & 52         \\
               &       &               & Query embellishment\change{~\cite{trippas2017people}}                 & 20         \\
               &       &                & Spells (query or query word)          & 2         \\
               \cmidrule{3-5}
               &       & \multirow{12}{*}{Intermediary}               & \change{(Requests)} Definition clarification \change{(i.e., requests more details about the information request)}         & 1        \\
               &       &                & Enquiry for further information           & 11         \\
               &       &                & Google query expansion suggestion          & 3         \\
               &       &                & Query refinement offer           & 57         \\
               &       &                & Query rephrase          & 12         \\
               &       &                & Requests more details about information request           & 5         \\
               &       &                & Query formulation for information found in document          & 1         \\
               &       &                & Asking what they \change{(i.e., the Seeker)} are looking for           & 2         \\
               &       &                & Within-Document search result entity lookup request           & 1         \\ 
               \bottomrule
\end{tabular}
}
\end{table}
}

{\renewcommand{\arraystretch}{.8}
\begin{table}[ht]
\centering
\caption{Result Presentation Codes (Intermediary)}
\label{tab:Result Presentation Codes (Intermediary)}
\adjustbox{max width=1\textwidth}{
\begin{tabular}{llllc}
\toprule
\textbf{Theme} & \textbf{Sub-theme}  & \textbf{Actor}  & \textbf{Code}      & \textbf{Frequency} \\
\midrule
\multirow{22}{*}{Task Level}   & \multirow{22}{*}{Results Presentation} & \multirow{22}{*}{Intermediary} & Source information                                                       & 8         \\
               &      &                  & Image overview on SERP                                                   & 2         \\
               &      &                  & Interpretation of photos                                                 & 1         \\
               &      &                 & Multi-document summary                                                   & 3         \\
               &      &                 & \begin{tabular}[c]{@{}l@{}}Paraphrasing from document which is \\ not in front of them\end{tabular}                 & 1        \\
               &      &                 & Scanning document with modification                                      & 51         \\
               &      &                & Scanning document without modification                                   & 79         \\
               &      &                & \begin{tabular}[c]{@{}l@{}}Scanning document without modification \\but with interpretation of photos \end{tabular} & 1         \\
               &      &                 & SERP Card                                                                & 16         \\
               &      &                 & SERP overview without modification                                       & 1         \\
               &      &                & SERP with modification                                                   & 19         \\
               &      &                 & SERP without modification                                                & 72         \\ 
                &     &                  & Within SERP search result                                                & 4         \\ 
                 &    &                  & Within-Document command response                                               & 1         \\ 
                 &    &                  & Within-Document search result                                             & 60         \\ 
                 &    &                  & \begin{tabular}[c]{@{}l@{}}Interpretation biased towards information\\ request or clarification given by the User  \end{tabular}                                             & 1         \\ 
                 &    &                  & Comparing results against each other                                                & 1         \\ 
                 &    &                  & Interpretation                                     & 22         \\ 
\bottomrule
\end{tabular}
}
\end{table}
}


{\renewcommand{\arraystretch}{.8}
\begin{table}[htp]
\centering
\caption{Search Assistance (Seeker and Intermediary)}
\label{my-label}
\adjustbox{max width=1\textwidth}{
\begin{tabular}{lllp{7cm}c}
\toprule
\textbf{Theme} & \textbf{Sub-theme}   & \textbf{Actor}        & \textbf{Code}                           & \textbf{Frequency} \\ 
\midrule
\multirow{10}{*}{Task Level}   & \multirow{10}{*}{Search Assistance} & \multirow{2}{*}{Seeker}       & \change{(Requests further search)} Recommendations                         & 1         \\
               &                      &        & Requests ``enough information" judgement & 1         \\
               \cmidrule{3-5}
               &                      &
               
               \multirow{8}{*}{Intermediary} & Asking about usefulness \change{(of presented result)}               & 4         \\
               &                      &  & Requests spelling                       & 2         \\
               &                      &  & Suggestion to move on                   & 2         \\
               &                      &  & Relevance judgement                     & 6         \\
               &                      &  & Suggestion to search more               & 1         \\
               &                      &  & Requests to access search engine        & 1   \\     
               &                      &  & Search suggestion based on info encountered in document & 1\\ 
               \bottomrule
\end{tabular}
}
\end{table}
}


{\renewcommand{\arraystretch}{.8}
\begin{table}[htp]
\centering
\smaller{}
\caption{Search Progression (Seeker)}
\label{tab:Search Progression or Meta-discussion (Seeker)}
\adjustbox{max width=1\textwidth}{
\begin{tabular}{llllc}
\toprule
\textbf{Theme} & \textbf{Sub-theme} & \textbf{Actor} & \textbf{Code}        & \textbf{Frequency} \\
\midrule
\multirow{3}{*}{Task Level}   & \multirow{3}{*}{Search Progression}  & \multirow{3}{*}{Seeker} & Enough information   & 6         \\
               &                    &  & Performance feedback & 18         \\
               &                    &  & Rejects \change{(suggestion from Intermediary)}              & 9         \\ 
               \bottomrule
\end{tabular}
}
\end{table}
}

\clearpage

\subsection{Theme 2: Discourse Level}
\label{appendix:Theme 2: Discourse Level}

{\renewcommand{\arraystretch}{.8}
\begin{table}[H]
\centering
\caption{Discourse Management (Seeker and Intermediary)}
\label{Discourse Management (Seeker and Intermediary)}
\adjustbox{max width=1\textwidth}{
\begin{tabular}{llllc}
\toprule
\textbf{Theme}           & \textbf{Sub-theme}            & \textbf{Actor}        & \textbf{Code}                        & \textbf{Frequency}  \\
\midrule
\multirow{10}{*}{Discourse Level} & \multirow{10}{*}{Discourse Management} & \multirow{5}{*}{Seeker}       & Asks to repeat              & 31 \\
                &                      &        & Asks to repeat first search result                    & 6 \\
                &                      &        & Asks to repeat Nth search result               & 1 \\
                &                      &        & Confirms                    & 114 \\
                &                      &        & Query repeat                & 14 \\
                \cmidrule{3-5}
                &                      & \multirow{5}{*}{Intermediary} & Asks to repeat              & 38 \\
                &                      &  & Checks navigational command & 13 \\
                &                      &  & Confirms                    & 46 \\
                &                      &  & Repeats                     & 12 \\
                &                      &  & Repeats the query back      & 9 \\ 
                \bottomrule
\end{tabular}
}
\end{table}
}

{\renewcommand{\arraystretch}{.8}
\begin{table}[H]
\centering
\smaller{}
\caption{Grounding (Seeker)}
\label{Grounding (Seeker)}
\begin{tabular}{llllc}
\toprule
\textbf{Theme}           & \textbf{Sub-theme} & \textbf{Actor}  & \textbf{Code}                    & \textbf{Frequency}  \\ \midrule
\multirow{2}{*}{Discourse Level} & \multirow{2}{*}{Grounding} & \multirow{2}{*}{Seeker} & Creating bigger picture & 1 \\
                &           &  & Interpretation          & 12 \\ 
                \bottomrule
\end{tabular}
\end{table}
}

{\renewcommand{\arraystretch}{.8}
\begin{table}[H]
\centering
\smaller{}
\caption{Navigation (Seeker)}
\label{Navigation (Seeker)}
\adjustbox{max width=1\textwidth}{
\begin{tabular}{llllc}
\toprule
\textbf{Theme}           & \textbf{Sub-theme}  & \textbf{Actor}  & \textbf{Code}                        & \textbf{Frequency}  \\ 
\midrule
\multirow{10}{*}{Discourse Level} & \multirow{10}{*}{Navigation} & \multirow{10}{*}{Seeker} & Access link within document & 1 \\
                &            &  & Access search engine    & 2 \\
                &            &  & Access source               & 29 \\
                &            &  & Access source (implicit)    & 2 \\
                &            &  & Between-document navigation & 1 \\
                &            &  & Is there more information   & 6 \\
                &            &  & Leave document              & 1 \\
                &            &  & Next                        & 3 \\
                &            &  & Read more from the document & 1 \\
                &            &  & Within-document command     & 3 \\ 
                \bottomrule
\end{tabular}
}
\end{table}
}


{\renewcommand{\arraystretch}{.8}
\begin{table}[H]
\centering

\caption{Visibility of System Status (Seeker and Intermediary)}
\label{Visibility of System Status (Seeker and Intermediary)}
\adjustbox{max width=1\textwidth}{
\begin{tabular}{lp{2cm}lp{5cm}c}
\toprule
\textbf{Theme}           & \textbf{Sub-theme}                   & \textbf{Actor}  & \textbf{Code}                           & \textbf{Frequency}  \\ 
\midrule
\multirow{7}{*}{Discourse Level} & \multirow{7}{2cm}{Visibility of system status} & \multirow{3}{*}{Seeker} & Access source feedback-request & 3 \\
                &                             &  & Feedback on what is happening  & 1 \\
                &                             &  & Results?                       & 10 \\
                \cmidrule{3-5}
          &                                   & \multirow{4}{*}{Intermediary}   & Feedback on what is happening & 13         \\
                &                             &    & Misheard                      & 1         \\
                &                             &    & Previously seen results       & 2         \\
                &                             &    & Wayfinding                    & 3         \\ 
                \bottomrule
\end{tabular}
}
\end{table}
}

\subsection{Theme 4: Other Level}
\label{subsec:Theme 4: Other Level}

{\renewcommand{\arraystretch}{.8}
\begin{table}[H]
\centering
\caption{Other Level (Seeker)}
\label{Other Level (Seeker)}
\adjustbox{max width=1\textwidth}{
\begin{tabular}{llp{11cm}c}
\toprule
\textbf{Theme} & \textbf{Actor} & \textbf{Code}                      & \textbf{Frequency} \\ 
\midrule
\multirow{6}{*}{Other Level}                      & \multirow{6}{*}{Seeker}         & Utter (``So I'm'' and ``Well so they are saying'') & 2         \\
 &                              & Provides information about the Search Engine (``So it's [a] search engine'') & 1         \\
   &                              & Asks if allowed to query embellish (``Actually can I add something else to that?'') & 1         \\
   &                              & Offers to spell (``[...] would you like me to spell it?'') & 1  
\\ \bottomrule
\end{tabular}
}
\end{table}
}

\section{Tasks and Backstories}
\label{sec:appendixB}


{\renewcommand{\arraystretch}{.8}
\begin{table}[H]
   \centering
 \smaller
 \caption{Example Search Tasks\label{tab:Example Search Tasks}}
\adjustbox{max width=1\textwidth}{
\begin{tabular}{p{1.5cm}p{11.5cm}}
\toprule
\textbf{Dimension}  & \textbf{Query and Example Backstory}                                                                                                                                                                                                                                                                                              \\ \midrule
\multirow{13}{*}{Remember}                  & \texttt{What river runs through Rome, Italy?}                                                                                                                                                                                                                                                                                     \\
                    & Many great cities have rivers running through them, as rivers facilitated trade and commerce as well as supplying fresh water to drink. You remember that Paris has the Seine, London has the Thames, but what does Rome have?                                                                                           \\ \cmidrule{2-2} 
                    & \texttt{What language do they speak in New Caledonia?}                                                                                                                \\
                    & You and your partner are thinking of places to go on holiday.  New Caledonia is an option, but you realize you don't know what language is spoken there and you decide to find out.                                                                                                                                      \\ \cmidrule{2-2} 
                    & \texttt{Where does cinnamon come from?}                                                                                                                                                                                                                                                                                           \\
                    & The other day you were eating some spiced biscuits from Europe, when it occurred to you that cinnamon probably isn't native to that part of the world.  You would like to know where it comes from.                                                                                                                      \\ \midrule
\multirow{12}{*}{Understand}          & \texttt{recycle, automobile tires                                }                                                                                                                                                                                                                                                                \\
                    & You need to buy new tires for your car, and the local dealer has offered to take the old ones for recycling.  You didn't know tires could be recycled and you wonder what new uses they are being put to.                                                                                                                \\ \cmidrule{2-2} 
                    & \texttt{Outsource job India             }                                                                                                                                                                                                                                                                                         \\
                    & A recent report on the radio quoted a politician as saying that one of the causes of rising unemployment in the U.S. was the outsourcing of jobs to India.  This has made you interested in finding out what jobs that used to be in the U.S. have been outsourced to India.                                             \\ \cmidrule{2-2} 
                    & \texttt{Marine Vegetation               }                                                                                                                                                                                                                                                                                         \\
                    & You recently heard a commercial about the health benefits of eating algae, seaweed and kelp. This made you interested in finding out about the positive uses of marine vegetation, both as a source of food, and as a potentially useful drug.                                                                           \\ \midrule
\multirow{16}{*}{Analyse}             & \texttt{Turkey Iraq Water}                                                                                                                                                                                                                                                                                                        \\
                    & Looking at a map, you realize that there are several rivers that commence in Turkey and then flow over the border into Iraq.  You wonder if Turkish river control projects, including dams and irrigation schemes, have affected Iraqi water resources.                                                                 \\ \cmidrule{2-2} 
                    & \texttt{Airport Security                }                                                                                                                                                                                                                                                                                         \\
                    & Every time you go through the security screening at an airport, you wonder whether it is making any difference. Find out how effective the many new measures (beyond just standard screening) at airports actually are, both for scrutinizing of passengers and their checked and carry-on baggage.                      \\ \cmidrule{2-2} 
                    & \texttt{per capita alcohol consumption  }                                                                                                                                                                                                                                                                                         \\
                    & You recently attended a big party and woke up with a hangover, and have decided to learn more about the average consumption of alcohol. You are particularly interested in any information that reports per capita consumption, and want to compare across groups, for example at the country, state, or province Level. \\
    \bottomrule
\end{tabular}
}
\end{table}
}

\end{document}